\documentclass[letterpaper, 10 pt, conference]{ieeeconf}  

\IEEEoverridecommandlockouts                              

\usepackage[dvipdfmx]{graphicx}
\usepackage{epsfig} 
\usepackage{times} 
\usepackage{amsmath} 
\usepackage{amssymb}  

\usepackage{algorithm}
\usepackage{algorithmicx}
\usepackage[noend]{algpseudocode}
\usepackage{color}
\usepackage{xcolor}
\usepackage{url}

\title  {\LARGE \bf
Online Multi-Agent Pickup and Delivery with Task Deadlines
}

\author{Hiroya Makino$^{1,*}$ and Seigo Ito$^{1}$
\thanks{$^{1}$ H. Makino and S. Ito are with Toyota Central R\&D Labs., Inc., 41-1, Yokomichi, Nagakute, Aichi, Japan.}
        \thanks{$^{*}$ Corresponding author. {\tt\small hirom@mosk.tytlabs.co.jp}}%
        }

\begin{document}

\maketitle
\thispagestyle{empty}
\pagestyle{empty}

\renewcommand{\thefootnote}{\fnsymbol{footnote}}
\footnote[0]{© 2024 IEEE.  Personal use of this material is permitted.  Permission from IEEE must be obtained for all other uses, in any current or future media, including reprinting/republishing this material for advertising or promotional purposes, creating new collective works, for resale or redistribution to servers or lists, or reuse of any copyrighted component of this work in other works.}
\renewcommand{\thefootnote}{\arabic{footnote}}

\begin{abstract}
Managing delivery deadlines in automated warehouses and factories is crucial for maintaining customer satisfaction and ensuring seamless production.
This study introduces the problem of online multi-agent pickup and delivery with task deadlines (MAPD-D), an advanced variant of the online MAPD problem incorporating delivery deadlines. 
In the MAPD problem, agents must manage a continuous stream of delivery tasks online. Tasks are added at any time. 
Agents must complete their tasks while avoiding collisions with each other. 
MAPD-D introduces a dynamic, deadline-driven approach that incorporates task deadlines, challenging the conventional MAPD frameworks.
To tackle MAPD-D, we propose a novel algorithm named deadline-aware token passing (D-TP). 
The D-TP algorithm calculates pickup deadlines and assigns tasks while balancing execution cost and deadline proximity. 
Additionally, we introduce the D-TP with task swaps (D-TPTS) method to further reduce task tardiness, enhancing flexibility and efficiency through task-swapping strategies. 
Numerical experiments were conducted in simulated warehouse environments to showcase the effectiveness of the proposed methods. 
Both D-TP and D-TPTS demonstrated significant reductions in task tardiness compared to existing methods. Our methods contribute to efficient operations in automated warehouses and factories with delivery deadlines.
\end{abstract}

\section{Introduction}
In the industrial automation landscape, the development of automated guided vehicles (AGVs) has revolutionized operational efficiencies. 
Enhancing multi-agent path finding (MAPF) to optimize the utilization of AGVs for more effective transportation solutions has been extensively researched \cite{stern2019, salzman2020}. 
These advancements have been integrated into various domains, including logistics automation \cite{wurman2008, honig2019}, traffic control systems \cite{dresner2008}, automated valet parking \cite{okoso2019}, airport surface operations \cite{li2019}, and video games \cite{silver2005}.

The multi-agent pickup and delivery (MAPD) problem is an extension of MAPF, wherein paths for multiple items are planned from pickup to delivery locations without collisions between agents \cite{ma2017, ma2019a, liu2019}. 
MAPD can be applied to various environments, including automated warehouses and factories \cite{ma2017, li2021, aryadi2023a}. 
In such settings, the importance of managing delivery deadlines cannot be overstated. 
Warehouses must tailor deadlines for individual orders to ensure customer satisfaction, whereas factories require timely deliveries to maintain seamless production. 
Therefore, satisfying the deadlines appropriately is essential for operational efficiency and economic success. 

\begin{figure}[!t]
  \centering
  \includegraphics[width=0.9\linewidth]{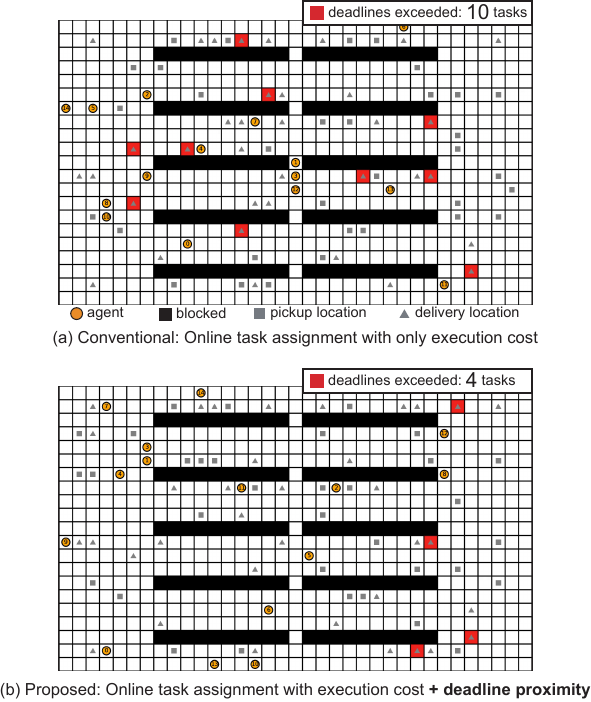}
  \caption{Simulation examples in a four-neighbor grid environment \cite{ma2017}.
  Unlike MAPD \cite{ma2017}, a deadline is set for each task in MAPD-D.
  Each agent searches for a path from the pickup to delivery locations for the assigned task.}
  \label{env_cover}
\end{figure}

\begin{figure}[!t]
  \centering
  \includegraphics[width=0.9\linewidth]{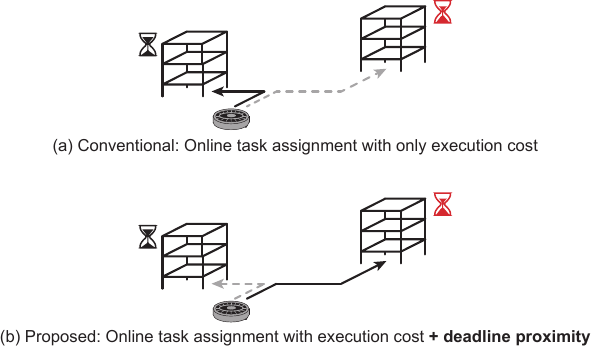}
  \caption{Comparison of task assignment methods: Our method assigns tasks by considering the execution cost and proximity of deadlines, thereby effectively reducing tardiness.}
  \label{env_cover_method}
\end{figure}

To consider deadlines in MAPD, we defined the online multi-agent pickup and delivery with task deadlines (MAPD-D) as a new problem (Fig. \ref{env_cover}). 
Existing studies \cite{wu2021, ramanathan2023} have explored MAPD with deadlines in an \textit{offline} setting, where all task information is provided in advance. 
However, to the best of our knowledge, no study has addressed deadline-aware MAPD in an \textit{online} setting where tasks may be added at any time. 
Hence, this study introduces new problem definitions for MAPD-D.

We propose the deadline-aware token passing (D-TP) algorithm to tackle MAPD-D. 
This algorithm is designed to calculate pickup deadlines and assign tasks while striking a balance between execution cost and deadline proximity (Fig. \ref{env_cover_method}). 
Additionally, we introduce the D-TP with task swaps (D-TPTS) method to reduce task tardiness. 
D-TPTS enhances flexibility and efficiency by employing task-swapping strategies among agents and within a single agent.

The primary contributions of this study can be summarized as follows:
\begin{itemize}
  \item A new problem (MAPD-D) is defined considering delivery deadlines and the possibility of adding tasks at any time.
  \item A method for solving MAPD-D by calculating pickup deadlines and deadline-aware task assignments is proposed. 
  \item Task-swapping methods are introduced to reduce task tardiness.
\end{itemize}
The remainder of this paper is structured as follows. 
Section \ref{sec:related_work} describes the related work on MAPD. 
Section \ref{sec:problem_definition} defines the MAPD-D problem, and Section \ref{sec:proposed_method} describes the proposed algorithm for solving this problem. 
Section \ref{sec:numerical_experiments} presents the numerical experiments performed to evaluate the proposed method. 
Finally, Section \ref{sec:conclusion} summarizes the study findings and concludes the paper.

\begin{table*}[!t]
  \vspace*{1.35mm}
  \centering
  \caption{Related work.}
  \footnotesize
  \label{related_work}
  \begin{tabular}{l|llll}
    \hline \hline                                                                           & Multi-task & Deadlines & Individual release times & Online \\ \hline
  Lifelong multi-agent path finding for online pickup and delivery tasks (MAPD) \cite{ma2017}& \checkmark      & $\times$         & \checkmark             & \checkmark                         \\ \hline
  Multi-agent path finding with deadlines (MAPF-DL) \cite{ma2018}                                & $\times$      & \checkmark         & $\times$             & $\times$                         \\ \hline
  Multi-robot path planning with due times (MRPP-DT) \cite{wang2022a}                                & $\times$      & \checkmark         & $\times$             & $\times$                         \\ \hline
  Deadline-aware multi-agent tour planning (DA-MATP) \cite{huang2023}                                & $\times$      & \checkmark         & $\times$             & $\times$                         \\ \hline
  Multi-agent pickup and delivery with task deadlines (MAPD-TD) \cite{wu2021}                   & \checkmark      & \checkmark         & $\times$             & $\times$                         \\ \hline
  Minimizing task tardiness for multi-agent pickup and delivery \cite{ramanathan2023}        & \checkmark      & \checkmark         & \checkmark             & $\times$                         \\ \hline
  Proposed (MAPD-D)                                                                 & \checkmark      & \checkmark         & \checkmark             & \checkmark \\ \hline                        
  \end{tabular}
\end{table*}

\section{Related Work}
\label{sec:related_work}
MAPF is the problem of moving multiple agents to their respective destination locations without collisions. 
The MAPF task ends when all agents reach their destinations. 
In contrast, MAPD requires agents to attend to a stream of delivery tasks \cite{ma2017, ma2019a}, wherein each delivery task is assigned a starting point (pickup location) and a destination point (delivery location). 
The system assigns these tasks to agents, which then move to the delivery location via the pickup location. 
The agents receive new tasks after reaching the delivery location. 
Ma et al. \cite{ma2017, ma2019a} described an \textit{online} version of MAPD, in which tasks can be added to the set at any time. 
Conversely, Liu et al. \cite{liu2019} discussed an \textit{offline} variant where tasks and their release times are generally predetermined.

To address online MAPD, Ma et al. \cite{ma2017} employed the token passing (TP) algorithm. 
Tokens are a type of shared memory that stores all agent paths as well as the task set and agent assignments. 
Agents sequentially access the token, are assigned tasks, and find paths without colliding with already reserved paths. 
The token is updated after identifying a path.

Several researchers have investigated MAPF and MAPD with deadlines. 
In this study, we classified related studies from the literature considering four perspectives (Table \ref{related_work}). 
``Multi-task'' indicates whether each agent is continuously assigned tasks;
``Deadlines'' denote whether each task is set with a deadline;
``Individual release times'' denote whether the release time of each task is the same; and 
``Online'' indicates whether tasks are added at any time. 
As summarized  in Table \ref{related_work}, Ma et al. \cite{ma2018}, Wang and Chen \cite{wang2022a}, and Huang et al. \cite{huang2023} considered deadlines in the context of MAPF. 
Wu et al. \cite{wu2021} and Ramanathan et al. \cite{ramanathan2023} introduced deadlines in MAPD.
However, the perspective of online tasks in MAPD was not considered in \cite{wu2021, ramanathan2023}. 
In this study, the proposed MAPD-D considers online tasks with deadlines, where tasks can be added at any time.

The dynamic vehicle routing problem with
pickups and deliveries (Dynamic VRPPD) also considers deadlines for online requests (tasks) \cite{berbeglia2010, pillac2013}.
Passengers' request time corresponds to deadlines, and one of the objective functions in Dynamic VRPPD is to minimize the sum of the difference between the request and pickup times (waiting time) \cite{alonso-mora2017}.
These settings are similar to MAPD-D; however, Dynamic VRPPD does not consider collisions between agents.

\section{Problem Definition}
\label{sec:problem_definition}
\subsection{MAPD-D problem}
In this sub-section, we describe the shared definitions of MAPD \cite{ma2017} and MAPD-D and define the tasks specific to MAPD-D.

An instance of MAPD and MAPD-D consists of $m$ agents in $A={a_1, \ldots, a_m}$, a connected simple undirected graph $G=(V,E)$, and a set of unexecuted tasks $\mathcal{T} = { \tau_1, \ldots, \tau_k }$.
Here, $V$ represents the vertices, $E$ denotes the edges, and $l_i(t) \in V$ indicates the location of agent $a_i$ at timestep $t$. 
A path is a sequence of vertices associated with timesteps, indicating the vertex at which the agent is located at each timestep.

Agents either remain at their current node $l_i(t)=l_i(t+1)$ or move to an adjacent node via an edge $\left(l_i(t), l_i(t+1) \right) \in E$ at each timestep. 
Agents must avoid collisions with each other. 
MAPD and MAPD-D define two types of collisions \cite{stern2019, ma2017}: 
(1) Vertex conflict, where two agents cannot occupy the same location at the same timestep, that is, for all agents $a_i, a_j \, (i \neq j)$ and all timesteps $t$, $l_i(t) \neq l_j(t)$ must hold; and 
(2) Swapping conflict, where two agents cannot move along the same edge in opposite directions at the same timestep, that is, for all agents $a_i, a_j \, (i \neq j)$ and all timesteps $t$, $l_i(t) \neq l_j(t+1)$ or $l_j(t) \neq l_i(t+1)$ must hold.

We further provide an extended definition of tasks in MAPD-D. 
Each task $\tau_j \in \mathcal{T}$ consists of a pickup location $v^{\mathrm{p}}_j \in V$ and delivery location $v^{\mathrm{d}}_j \in V$.
In an extended definition for MAPD-D, each task additionally includes a delivery deadline $d^\mathrm{d}_j$. 
$d^\mathrm{d}_j$ is not a tight bound, and agents are still expected to take on new tasks and proceed even if a task's deadline has passed. 
When an agent assigned to a task reaches the delivery location via the pickup location, the task is completed, and its completion time is denoted as $c_j$. 
Once a task is completed, the agent is assigned a new task. 
New tasks can be added to the task set $\mathcal{T}$ at each timestep. 

The objective of MAPD is to identify paths that execute all tasks promptly, whereas MAPD-D aims to minimize task tardiness. 
We define the tardiness for the $j$-th task as $\epsilon_j = \max(0, c_j - d^\mathrm{d}_j)$.
For each task $\tau_j$, $\epsilon_j = 0$ indicates the success of the task, whereas $\epsilon_j > 0$ indicates task failure. 
The objective function of MAPD-D is selected from the following two options:
\begin{itemize}
  \item Minimizing the number of task failures \cite{wu2021, ma2018, huang2023}: \begin{align} \min \, \sum_{1 \leq j \leq k} U(\epsilon_j), \label{eq:obj1} \end{align} where $U(\cdot)$ is a unit step function
  \footnote{The unit step function.

    \begin{minipage}{\linewidth}
    \begin{align*}
    U(x) &= \begin{cases} 0 & \text{if } x \leq 0 \\ 1 & \text{if } x > 0. \end{cases}
    \end{align*}
    \end{minipage}
  }. 
  \item Minimizing the cumulative tardiness \cite{ramanathan2023, wang2022a}: \begin{align} \min \, \sum_{1 \leq j \leq k} \epsilon_j. \label{eq:obj2} \end{align} 
\end{itemize}
In this study, we used the cumulative tardiness \eqref{eq:obj2} as the objective function because it provides a more detailed evaluation of tardiness compared with the objective function \eqref{eq:obj1}, which only considers success or failure.

\subsection{Well-formed MAPD-D Instances}
\label{sec:well_formed}
Ma et al. \cite{ma2017} reported that online MAPD instances are solvable if they are well-formed. 
They introduced locations referred to as endpoints, where agents can remain without blocking other agents.
Endpoints include all pickup and delivery locations along with the initial positions and designated parking locations. 
The pickup and delivery locations are referred to as task endpoints, whereas the other locations serve as non-task endpoints. 
An MAPD instance is well-formed if and only if (a) the number of tasks is finite, (b) the number of agents does not exceed the number of non-task endpoints, and (c) a path exists between any two endpoints without traversing others.

In MAPD-D, only the task deadline is introduced, while other task settings and environments are the same as in MAPD. 
Therefore, the definition of well-formed MAPD-D instances is the same as that for MAPD.

\section{Proposed Method}
\label{sec:proposed_method}
This section outlines the D-TP algorithm employed to address the MAPD-D problem, which is an extension of the existing TP method. 
Typically, TP assigns tasks to agents solely based on the execution cost. 
To reduce task tardiness, we introduce enhancements in two key areas: the calculation of pickup deadlines and deadline-aware task assignments. 
Additionally, we present D-TPTS, which improves flexibility and efficiency through task-swapping strategies.

\subsection{D-TP}
\subsubsection{Calculation of Pickup Deadlines}
The pickup deadline is calculated based on the delivery deadline when a new task $\tau_j$ is added. 
This deadline is not a tight bound but rather a reference value used for cost calculation in subsequent task assignments. 

We prepared a dummy agent and implemented it using prioritized path planning \cite{liu2019}. 
D-TP calculates the path for the dummy agent to depart from the delivery location $v^{\mathrm{d}}_j$ at time $d^\mathrm{d}_j$ and move toward the pickup location $v^{\mathrm{p}}_j$ by reversing the timesteps. 
During the path calculation, the order of the pickup and delivery locations is reversed because the time required for transportation can vary depending on the paths of other agents in the environment. 
The proposed method searches for a path from the delivery location to the pickup location by reversing the order of time, thus calculating the latest possible path (dummy path) that meets the delivery deadline. 
The pickup deadline $d^\mathrm{p}_j$ represents the time obtained by subtracting the length (timesteps) of the dummy path from the delivery deadline $d^\mathrm{d}_j$.

\subsubsection{Deadline-aware Task Assignment}
The disparity between the calculated pickup deadline $d^\mathrm{p}_j$ and current time $t$ indicates the temporal margin of the deadline. 
In TP, the system assigns tasks to minimize the execution cost at the moment of assignment. 
Contrastingly, in D-TP, the system assigns tasks to agents in a manner that minimizes the weighted sum of the execution cost and temporal margin relative to the deadline.
\begin{align}
\label{eq:obj}
\underset{\tau_j \in \mathcal{T}^\prime} {\operatorname{argmin}} \left(\alpha \cdot (d^\mathrm{p}_j - t) + (1 - \alpha) \cdot h \left(loc(a_i), v^{\mathrm{p}}_j \right) \right),
\end{align}
where $0\leq \alpha \leq 1$ and $\mathcal{T}^\prime$ denotes the set of tasks that can be assigned to agent $a_i$.
The first term $(d^\mathrm{p}_j - t)$ denotes the temporal margin for the deadline of task $\tau_j$; the second term $h( loc(a_i), v^{\mathrm{p}}_j)$ indicates the h-value from the current location of agent $a_i$ to the pickup location of the task, which represents the execution cost; and parameter $\alpha$ indicates the weight for the urgency of the pickup deadline. 
When $\alpha=0$, it is equivalent to the existing method \cite{ma2017} that does not consider the deadline.

\subsubsection{Algorithm}
Algorithm \ref{algorithm_tpd} provides the pseudocode for D-TP; parts that differ from TP \cite{ma2017} are indicated in blue. 
In lines 17--20, we define the \textsc{UpdatePickupDeadline} function.
In line 19, a dummy agent $a_{dmy}$ is prepared to calculate the dummy path from the delivery location $v^{\mathrm{d}}_j$ to the pickup location $v^{\mathrm{p}}_j$ using the reversed-path finding function $\mathcal{RP}(dummy, v^{\mathrm{d}}_j, v^{\mathrm{p}}_j, token, d^\mathrm{d}_j)$. 
In line 20, the pickup deadline $d^\mathrm{p}_j$ is calculated by subtracting the length of the dummy path.
Here, $|\mathrm{path}|$ represents the length of the path, that is, the number of steps required to move.

In line 1, the token is initialized with trivial paths where all agents remain in their initial locations.
At each timestep, the system adds all new tasks to the task set $\mathcal{T}$ (line 3). 
Subsequently, in line 4, the pickup deadline is calculated for the newly added tasks. 
Lines 5--15 handle the task assignment process.
If one or more tasks are assignable, the system assigns a task considering both the execution cost and the margin until the pickup deadline $d^\mathrm{p}_j$. 
The path from $v^{\mathrm{p}}_{j}$ to $v^{\mathrm{d}}_{j}$ is calculated by the function $\mathcal{P}(a_i, v^{\mathrm{p}}_{j}, v^{\mathrm{d}}_{j}, token, t)$.
In cases where no tasks are assignable, the system handles deadlock resolution or maintains the current position of agents, as outlined in \cite{ma2017}. 
If new task assignments overwrite the dummy paths, the pickup deadline is recalculated (line 15). 
Finally, agents proceed along their paths in the token (line 16).

\begin{figure}[!t]
  \begin{algorithm}[H]
    \caption{Token passing for tasks with deadlines (TP-D)}
    \label{algorithm_tpd}
      \begin{algorithmic}[1]
        \footnotesize
        \State Initialize $token$ with the (trivial) path $[loc(a_i)]$ for each agent $a_i$
        \While{true}
          \State Add all new tasks, if any, to the task set $\mathcal{T}$
          \color{blue}\State \Call{UpdatePickupDeadline}{new tasks, $token$} \color{black}
          \While{agent $a_i$ that requests $token$ exists}
            \State $\mathcal{T}^\prime \leftarrow \{ \tau_j \in \mathcal{T} \mid \text{no other path in }token \text{ ends in } v^{\mathrm{p}}_j \text{ or } v^{\mathrm{d}}_j\}$
            \If{$\mathcal{T}^\prime \neq \phi$}
              \State $t \leftarrow $ current timestep
              \color{blue} \State $\tau_{j^*}$ is calculated by \eqref{eq:obj} 
              \color{black} \State Assign $a_i$ to $\tau_{j^*}$
              \State Remove $\tau_{j^*}$ from $\mathcal{T}$
              \State Update $a_i$'s path in $token$ with $\mathcal{P}(a_i, v^{\mathrm{p}}_{j^*}, v^{\mathrm{d}}_{j^*}, token, t)$
            \Else
              \State Remove deadlock or stay
            \EndIf
            \color{blue}\State \Call{UpdatePickupDeadline}{$\mathcal{T}$ whose dummy path is overwritten, $token$} \color{black}
          \EndWhile
          \State All agents move along their paths in token for one timestep
        \EndWhile
        \color{blue} \Function {UpdatePickupDeadline}{$tasks, token$}
          \For{$\tau_j \in tasks$}
            \State Update $\tau_j$'s dummy path with $\mathcal{RP}(a_{dmy}, v^{\mathrm{d}}_j, v^{\mathrm{p}}_j, token, d^\mathrm{d}_j)$
            \State Update $\tau_j$'s pickup deadline $d^\mathrm{p}_j$ with $d^\mathrm{d}_j - |\mathrm{dummy\ path}|$
          \EndFor
        \EndFunction \color{black}
      \end{algorithmic}
  \end{algorithm}
\end{figure}

\subsection{D-TPTS}
In this section, we introduce two methods for task swapping that incorporate deadlines in MAPD-D: \textit{Task swapping among agents} and \textit{task switching}. 
\textit{Task swapping among agents} involves swapping tasks between agents, while \textit{task switching} focuses on swapping tasks within a single agent.

\subsubsection{Task Swapping Among Agents}
Ma et al. \cite{ma2017} proposed the token passing with task swaps (TPTS) algorithm as a solution to MAPD. 
In TP with task swapping, agents can be assigned not only ``unassigned'' tasks but also ``tasks assigned to a different agent but not yet picked up.''
This flexibility can be advantageous, particularly when one agent can pick up a task faster than another. 
In such cases, the system reassigns the task from one agent to another, allowing for more efficient task completion.

Contrary to TPTS, which focuses solely on execution cost considerations, the proposed method modifies this approach to incorporate a weighted sum of the execution cost and temporal margin for deadlines, as expressed in \eqref{eq:obj}.

\subsubsection{Task Switching}
In this approach, the agent is allowed to abandon its current task and undertake a more urgent task if a task with higher urgency appears closer when an agent is en route to the pickup location.
We anticipate that task switching can reduce tardiness by prioritizing tasks with higher urgency.

An agent will abandon its current task if both of the following conditions are met:
\begin{itemize}
  \item The urgency of the new task is higher than that of the current task.
  \item The execution cost of the new task is lower than that of the current task.
\end{itemize}
Specifically, the following inequalities should hold simultaneously:
\begin{align}
  &d^\mathrm{p}_\mathrm{new} < d^\mathrm{p}_\mathrm{cur}, \\
  &h \left(loc(a_i), v^{\mathrm{p}}_\mathrm{new} \right) < h \left(loc(a_i), v^{\mathrm{p}}_\mathrm{cur} \right),
\end{align}
where $\mathrm{cur}$ denotes the index of the current task of agent $a_i$ and $\mathrm{new}$ represents the index of the new task.

\subsubsection{Algorithm}
Algorithm \ref{algorithm_tptsd} provides the pseudocode for D-TPTS. 
The overall procedure mirrors that of TPTS \cite{ma2017}, with differences highlighted in blue. 
Lines 5--10 implement task switching. 
When an agent is en route to the pickup locations, it abandons its current task and accepts a different task if the new task has an earlier pickup deadline and a lower execution cost than the current task. 
Task swapping is performed in the function \textsc{GetTask} (lines 15--32). 
As indicated in line 19, the system considers both the temporal margin of the pickup deadline and the weighted sum of the task execution cost.

\begin{figure}[!t]
  \begin{algorithm}[H]
    \caption{Deadline-aware token passing with task swaps (D-TPTS)}
    \label{algorithm_tptsd}
      \begin{algorithmic}[1]
        \footnotesize
        \State Initialize $token$ with the (trivial) path $[loc(a_i)]$ for each agent $a_i$
        \While{true}
          \State Add all new tasks, if any, to the task set $\mathcal{T}$
          \color{blue}\State \Call{UpdatePickupDeadline}{new tasks, $token$} \color{black}
          \color{blue} \For{$\tau_j \in$ new tasks}
            \For{agent $a_i $ that is moving to the pickup location}
              \State $\tau_{j^\prime} \leftarrow$ task that $a_i$ is executing 
              \If{$d^\mathrm{p}_j < d^\mathrm{p}_{j^\prime}$ and $h (loc(a_i), v^{\mathrm{p}}_j ) < h (loc(a_i), v^{\mathrm{p}}_{j^\prime} )$}
                \State Unassign $a_i^\prime$ from $\tau_{j^\prime}$
                \State Remove $a_i^\prime$'s path from $token$
              \EndIf
            \EndFor
          \EndFor \color{black}
          \While{agent $a_i$ that requests $token$ exists}
          \State \Call{GetTask}{$a_i, token$}
          \EndWhile
          \State All agents move along their paths in token for one timestep
          \State Remove tasks from $\mathcal{T}$ when agents start to execute them
          \EndWhile
        \Function {GetTask}{$a_i, token$}
          \State $\mathcal{T}^\prime \leftarrow \{ \tau_j \in \mathcal{T} \mid \text{no other path in }token \text{ ends in } v^{\mathrm{p}}_j \text{ or } v^{\mathrm{d}}_j\}$ 
          \While{$\mathcal{T}^\prime \neq \phi$}
            \State $t \leftarrow $ current timestep
            \color{blue} \State $\tau_{j^*}$ is calculated by \eqref{eq:obj} \color{black}
            \State Remove $\tau$ from $\mathcal{T}$
            \If{no agent is assigned to $\tau_{j^*}$}
              \State Assign $a_i$ to $\tau_{j^*}$
              \State Update $a_i$'s path in $token$ with $\mathcal{P}(a_i, v^{\mathrm{p}}_{j^*}, v^{\mathrm{d}}_{j^*}, token, t)$
            \Else
              \State $a_i^\prime \leftarrow$ agent that is assigned to $\tau_{j^*}$
              \If{$a_i$ reaches $v^{\mathrm{p}}_{j^*}$ before $a_i^\prime$}
                \State Unassign $a_i^\prime$ from $\tau_{j^*}$ and assign $a_i$ to $\tau_{j^*}$
                \State Remove $a_i^\prime$'s path from $token$
                \State Break                
              \EndIf
            \EndIf
          \EndWhile
          \If{no task is assigned to $a_i$}
            \State Remove deadlock or stay
          \EndIf
          \color{blue}\State \Call{UpdatePickupDeadline}{$\mathcal{T}$ whose dummy path is overwritten, $token$} \color{black}
        \EndFunction
      \end{algorithmic}
  \end{algorithm}
\end{figure}

\subsection{Completeness}
In well-formed MAPD instances, a path exists among the current location of an agent,  pickup location, and delivery location.
Additionally, the number of tasks is finite.
TP eventually assigns a task to each agent, and each agent can execute the task.
Therefore, TP solves any well-formed instances.

As discussed in Section \ref{sec:well_formed}, well-formed MAPD-D instances also satisfy the conditions for being well-formed in MAPD.
The difference between TP and D-TP is only in the cost calculation, and all other parts are the same.
Therefore, D-TP can solve well-formed instances, just like TP in MAPD.

We also introduce task switching in D-TPTS.
An agent may abandon its current task when a highly urgent task appears; however, the agent always finishes its current task within a finite time.
Therefore, D-TPTS can solve well-formed instances in the same way that TPTS solves them in MAPD.

\section{Numerical Experiments}
\label{sec:numerical_experiments}
This section outlines the numerical experiments conducted to compare the existing method (TP) with the proposed algorithms (D-TP and D-TPTS). 
Our primary focus is to evaluate the effectiveness of the proposed algorithms in reducing tardiness in an online setting.

\subsection{Evaluation of Task Tardiness}
\begin{figure}[!t]
    \vspace*{1.3mm}
  \centering
  \includegraphics[width=0.95\linewidth]{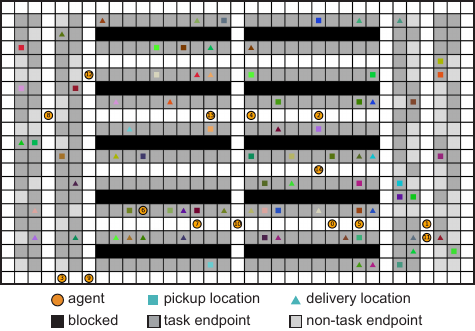}
  \caption{Four-neighbor grid environment \cite{ma2017}. A pair of pickup and delivery locations with the same color represents a single task. In MAPD-D, each delivery location is assigned a deadline.}
  \label{exp_env}
\end{figure}

Numerical experiments were conducted in a grid environment, representing an automated warehouse, as illustrated in Fig. \ref{exp_env}. 
We generated 151 tasks by selecting pickup and delivery locations uniformly at random from the task endpoints, ensuring no duplication. 
Each agent moved to the delivery location via the pickup location for the assigned task.
We varied parameters such as task-release times and deadline duration after task release to examine their significant impact during the experiments (see Table \ref{task_setting}). 
The experiment involved 15 agents, with their initial positions randomly selected from the non-task endpoints.

We also examined an offline setting where tasks and their release times are predetermined, contrary to the online setting where tasks can be added at any time. 
In an offline setting, tasks can be allocated with foresight to accommodate future tasks and preemptively moved in anticipation of release times. 
Generally, offline methods tend to yield solutions closer to the optimal solution compared to online methods. 
Therefore, we referred to the method by Ramanathan et al. \cite{ramanathan2023}, who regarded deadlines in offline tasks as ideal benchmarks. 
They employed a method that sorted tasks based on deadlines and assigned tasks to agents with lower execution costs.

\begin{figure*}[!t]
  \vspace*{1.35mm}
  \centering
  \includegraphics[width=0.8\linewidth]{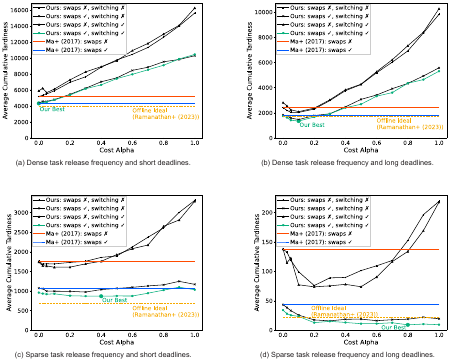}
  \caption{Comparisons of cumulative tardiness.}
  \label{plot_task_swaps}
\end{figure*}

Fig. \ref{plot_task_swaps} illustrates the results of the proposed method alongside ideal values in the offline setting. 
The horizontal axis represents the weight $\alpha$ used during task assignment, where smaller values prioritize execution cost and larger values prioritize deadlines. 
When $\alpha=0$ and task switching is not employed, our proposed new task swapping method is equivalent to TP and TPTS \cite{ma2017}. 
The vertical axis indicates the total tardiness for each task, averaged across the results of 30 experiments. 
Our analysis demonstrates that the cumulative tardiness varies depending on the value of $\alpha$ and the presence of task exchanges, regardless of the release frequency and deadlines.

\begin{table}[!t]
  \centering
  \caption{Settings of task release frequency and deadlines.}
  \footnotesize
  \label{task_setting}
  \begin{tabular}{l|ll}
    \hline \hline
    Task-release times & Dense [0, 300] & Sparse [0, 500] \\ \hline
    Deadline duration after release  & Short [20, 80] & Long [60, 120] \\ \hline 
    \multicolumn{3}{l}{*Values are randomly selected from a uniform distribution} \\ 
    \multicolumn{3}{l}{\phantom{*}in [min, max].} \\ 
  \end{tabular}
\end{table}

\subsection{Discussion}
We begin by examining the variations in tardiness based on task-release frequency and deadline length. 
In Fig. \ref{plot_task_swaps}, both the proposed and conventional methods exhibit notable trends in tardiness. 
Tardiness increases when tasks are released frequently and when deadlines are short. 
Densely released tasks leave agents with limited spare time, leading to a buildup of unexecuted tasks and an increase in cumulative tardiness. 
Similarly, shorter deadlines elevate the likelihood of tasks missing their deadlines, further contributing to cumulative tardiness.

Next, we explore the weight $\alpha$ in the proposed method. 
We compare the disparities in cumulative tardiness caused by varying $\alpha$ when neither task swapping nor task switching is implemented. 
Minimum tardiness is observed for dense task releases and short deadlines (Fig. \ref{plot_task_swaps}(a)) at $\alpha=0.0$, for dense releases and long deadlines (Fig. \ref{plot_task_swaps}(b)) at $\alpha=0.1$, for sparse releases and short deadlines (Fig. \ref{plot_task_swaps}(c)) at $\alpha=0.025$, and for sparse releases and long deadlines (Fig. \ref{plot_task_swaps}(d)) at $\alpha=0.2$. 
When tasks are released frequently and deadlines are short, prioritizing tasks with lower execution costs is crucial for timely management. 
However, in scenarios with less frequent task releases or longer deadlines, there is some flexibility to consider deadlines. 
While TP and TPTS \cite{ma2017} solely considered execution costs ($\alpha=0$), the proposed method integrates temporal margin for deadlines along with execution costs, resulting in reduced tardiness. 
Nonetheless, excessively prioritizing the temporal margin of deadlines may escalate the execution costs of each task.
This can overwhelm agents and increase cumulative tardiness. 
Hence, adjusting the value of $\alpha$ based on the situation is imperative.

In all experiments, the lowest tardiness was achieved when both task swapping and task switching were implemented. 
For example, as shown in Fig. \ref{plot_task_swaps}(b), implementing only task swapping reduced cumulative tardiness by $569.4$, whereas implementing both task swapping and task switching further reduced tardiness by an additional $116.7$. 
Task switching facilitates task reassignment when more urgent tasks are added. 
Additionally, task swapping enables task exchanges between agents, further reducing cumulative tardiness.

However, implementing only task switching may elevate cumulative tardiness compared to not implementing any strategy, particularly  when task releases are frequent (Fig. \ref{plot_task_swaps}(a) and \ref{plot_task_swaps}(b)). 
This aligns with the discussion on the value of $\alpha$; in scenarios with frequent task releases, minimizing execution costs outweighs considering task urgency. 
Task reassignment based on urgency through task switching increases execution costs, consequently amplifying tardiness.

Finally, we compare the proposed method (online) with the ideal values (offline). 
In most cases, the tardiness of the proposed method was equivalent to or worse than the ideal values (Fig. \ref{plot_task_swaps}(a), (c), and (d)). 
However, in scenarios with dense tasks and long deadlines, the proposed method outperformed the ideal values (Fig. \ref{plot_task_swaps}(b)). 
Ramanathan et al. \cite{ramanathan2023} sorted tasks in advance based on deadlines and assigned them to agents with lower execution costs.
This indicates that they prioritized deadlines over execution costs.
They noted that their method excelled with extremely short deadlines, exhibiting less tardiness than the proposed method in settings with frequent releases and short deadlines. 
However, maintaining a balance between execution costs and tardiness becomes crucial when handling numerous tasks and longer deadlines. 
The effectiveness of the proposed method is evidenced in such scenarios despite operating online.

\section{Conclusions}
\label{sec:conclusion}
This study addresses task deadlines by introducing a modified version of the MAPD problem, termed online MAPD-D. 
In online MAPD-D, tasks can be added at any time and assigned deadlines. 
To address MAPD-D, we propose two algorithms: D-TP and D-TPTS.
D-TP allocates tasks by considering their pickup deadlines along with execution costs, while D-TPTS facilitates task exchanges among agents and within a single agent. 
The numerical experiments demonstrated that both D-TP and D-TPTS effectively reduced task tardiness compared with existing methods.

These experiments were conducted in a $35 \times 21$ grid environment; however, our ability to solve MAPD-D in larger environments is limited owing to computational constraints. 
In the future, exploring the development of decentralized algorithms could enable the solution of large-scale MAPD-D. 
Additionally, algorithms should be devised to handle more realistic scenarios, such as paths being obstructed by uncertain obstacles \cite{shofer2023}.
Furthermore, optimizing the parameter $\alpha$ in \eqref{eq:obj} is necessary. While $\alpha$ can be tuned based on past trends, it is challenging to tune $\alpha$ for future trends in online task settings.  

\section*{Acknowledgments}
We thank Kenji Ito, Keisuke Otaki, and Yasuhiro Yogo for their insightful inputs and discussions.


\end{document}